\documentclass[12pt]{article}
\usepackage[T1]{fontenc}
\usepackage[utf8]{inputenc}
\usepackage[margin=1in]{geometry}
\usepackage[affil-it]{authblk} 
\usepackage{etoolbox}
\usepackage{lmodern}
\usepackage{cite}
\usepackage{changes}
\usepackage{microtype}
\usepackage{booktabs}
\usepackage{amssymb}
\usepackage{float}
\DeclareUnicodeCharacter{030C}{\v{c}}

\title{Quantum Information Processing With Integrated Silicon Carbide Photonics}

\def\correspondingauthor{\footnote{Corresponding author: smajety@ucdavis.edu}}

\author[1]{Sridhar Majety \correspondingauthor{}}
\author[1]{Pranta Saha}
\author[1,2]{Victoria A. Norman}
\author[1]{Marina Radulaski}
\affil[1]{Department of Electrical and Computer Engineering, University of California, Davis, CA 95616, USA}
\affil[2]{Department of Physics, University of California, Davis, CA 95616, USA}

\date{\vspace{-2em}}

\usepackage{graphicx}
\usepackage{amsmath}
\usepackage{braket}
\usepackage{titlesec}
\usepackage{siunitx}
\begin{document}

\maketitle
\section*{Abstract}
Color centers in wide band gap semiconductors are prominent candidates for solid-state quantum technologies due to their attractive properties including optical interfacing, long coherence times, spin-photon and spin-spin entanglement, as well as the potential for scalability.  Silicon carbide color centers integrated into photonic devices span a wide range of applications in quantum information processing, in a material platform with quantum-grade wafer availability and advanced processing capabilities. Recent progress in emitter generation and characterization, nanofabrication, device design, and quantum optical studies have amplified the scientific interest in this platform. We provide a conceptual and quantitative analysis of the role of silicon carbide integrated photonics in three key application areas: quantum networking, simulation, and computing. 

\section{Introduction}
Quantum mechanical effects such as superposition and entanglement open the doors to novel quantum information processing (QIP) technologies in communication, computation, sensing, and metrology, that are hard or impossible to build using conventional classical technologies. Among the various implementations with superconducting qubits, trapped ions, and neutral atoms, the solid-state systems with optically addressable spin states (quantum dots, crystal defects) have shown the most promise in distributed QIP owing to the ability to create entanglement between the spin degree of freedom and a photon(s), serving as stationary and flying qubit(s), respectively \cite{hensen2015loophole}. Having photons as couriers of quantum information is beneficial, as they are unaffected by thermal noise and can travel long distances via fiber-optic cable without interacting with each other. Flying qubits from separate systems can be transported to a common location to perform entanglement swapping, thus creating remote entanglement between systems without their direct interaction. Moreover, solid-state platforms are scalable and offer a chip-scale integration. 

For the purposes of QIP, one of the primary requirements is the ability of a single photon emitter (SPE) to generate indistinguishable photons that can be mutually entangled. There have been notable strides using quantum dots for this purpose \cite{stockill2017phase, tomm2021bright}, however, the significant variability in the emission wavelengths (with the quantum dot size, charge, and temperature) has been challenging to overcome in a scalable fashion. On the other hand, the color centers, which are point defects in wide bandgap semiconductors that behave like quasi-atoms, have nearly identical and well reproducible properties.

Color center platforms have not yet realized their full QIP potential. Several remarkable demonstrations have been made in this regard, such as spin-photon entanglement \cite{bernien2013heralded}, remote entanglement of solid-state spins separated by 1.3 km \cite{hensen2015loophole}, a multinode quantum network \cite{pompili2021realization}, implementation of fault-tolerant operation of a logical qubit \cite{abobeih2021fault}, memory-enhanced quantum communication \cite{bhaskar2020experimental}, and a quantum register that uses nine nuclear spins coupled to a single electron spin, with quantum memory nearing one minute \cite{bradley2019ten}. Most of these demonstrations were made using the  nitrogen-vacancy (NV) center in diamond, which has defined the color center field that is now populated by a variety of quantum emitters with complementary or advanced properties.

There are several challenges to overcome in order to scale the experimental demonstrations to large-scale implementations and to harness the full potential of quantum technologies. For applications in quantum networks, further research is required to maximize the collection of emitted photons into fibers, waveguides, and minimize the optical losses in the fiber-optic cable and operational errors. The field of quantum computation can be further boosted by developing quantum memories with long coherence times (few seconds or higher), high-fidelity single-shot readouts of spin states, and high fidelity quantum gates. Color centers emerging in diamond and silicon carbide (SiC) are optically addressable with properties like long spin coherence times, nearly identical emission wavelengths. These properties coupled with methods like defect engineering, nanophotonic integration, and better error correction protocols can overcome some of the existing challenges. 

There have been multiple useful review articles on silicon carbide and color center photonics \cite{norman2021novel, son2020developing, castelletto2020silicon, zhang2020material, bathen2021manipulating, lukin2020integrated} which guide the reader through the variety of emitter properties, devices, and techniques. In this perspective paper, we fill the gap in understanding of the concepts of the next generation of integrated optical quantum hardware. We believe our perspectives will inform scientists working on tangential areas of semiconductor defects and quantum information processing interested in the opportunities in integrated SiC quantum photonics. We open by discussing quantum optical properties of SiC color centers and suitability for quantum hardware, followed by a review of their integration into photonic devices. We then outline three key applications in quantum networking, simulation, and computing and discuss their implementation prospects with integrated color center photonics in SiC. 

\section{QIP with color centers in SiC}
In this section, we discuss methods for characterizing quantum optical properties of SiC color centers, review the set of characterized emitters in the field, and discuss how their properties are suitable for quantum information processing.

\subsection{Color centers in SiC}
\begin{figure}[htbp]
    \centering
    \includegraphics[scale=0.5]{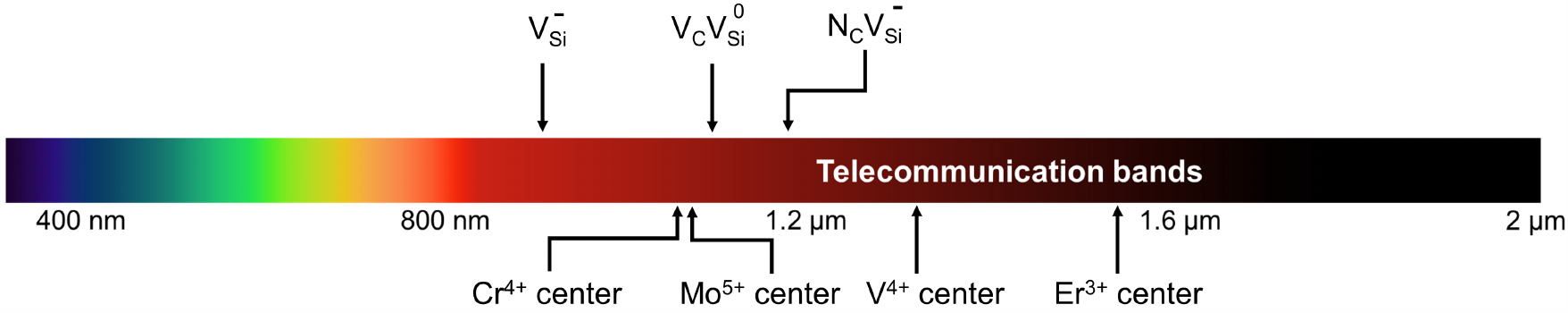}
    \caption{Emission wavelengths of color centers in 4H-SiC on the electromagnetic spectrum.  
    }
    \label{fig:fig1}
\end{figure}

In this section, we discuss the color centers in the most common polytypes of SiC i.e., 3C-SiC (cubic lattice), 4H- and 6H-SiC (hexagonal lattice). SiC hosts a variety of color centers like divacancy, silicon vacancy, nitrogen vacancy, carbon antisite-vacancy pair, and transition metal defects (chromium, vanadium, molybdenum, tungsten) with emission wavelengths in the near infrared region as shown in Figure \ref{fig:fig1}. Each defect in the lattice can exist in multiple orientations due to the presence of inequivalent lattice sites: hexagonal-like (h) and cubic-like (k). The defects in SiC exist naturally in as-grown material or can be \textit{in situ} doped during the crystal growth process. However, most of the studied color centers are generated through implantation where projectiles (electrons, protons, neutrons, ions) with sufficient energy are penetrated through the material surface. The implantation is followed by annealing in vacuum or inert atmosphere to activate the implants. This also helps repair the crystal lattice damage caused during implantation, though not completely \cite{gurfinkel2009ion}. The residual damage can create pathways for decoherence of the electron spin, which is a disadvantage while building QIP hardware. This can partly be overcome by using lighter projectiles like electrons, protons (over heavier ions) on \textit{in situ} doped substrates. In diamond, low energy ions were used to create shallow defects followed by growth of an high-quality epitaxial layer \cite{staudacher2012enhancing, rugar2020generation}, but this is not applicable to SiC due to large epitaxial growth temperatures.

The neutral divacancy defect (V\textsubscript{Si}V\textsubscript{C}$^0$) in SiC consists of a pair of neighboring silicon and carbon vacancies. The different possible orientations for each of the vacancies in 4H-SiC result in the photoluminescence of the divacancy to have six sharp lines (PL1-PL6) in the wavelength range of 1100 nm (PL1, PL2 - axial divacancies and PL3, PL4 - basal divacancies) \cite{koehl2011room}. Divacancy is a spin-1 system (like diamond NV center), but has advantageous properties like DWF of 5\% \cite{crook2020purcell}, highly-spin dependent optical fine structure (high-fidelity initialization and single shot readout) \cite{christle2017isolated}, and spin-coherence time of up to 5 seconds\cite{anderson2021five}. At the emission wavelengths of the divacancies, the fidelity of quantum frequency conversion to a telecommunications band is higher than for NV centers in diamond \cite{pelc2010influence}. 

The negatively charged silicon vacancy (V\textsubscript{Si}$^{^-}$) in SiC is formed at the site of a missing Si atom in the lattice and is a spin-3/2 defect. In 4H-SiC, V\textsubscript{Si}$^{^-}$ has two inequivalent lattice sites (h, k), resulting in emission spectrum consisting of two ZPLs i.e., V1 (861.6 nm) and V2 (916.5 nm) \cite{sorman2000silicon, janzen2009silicon}. Similarly in 6H-SiC, V\textsubscript{Si}$^{^-}$ has three inequivalent lattice sites (h, k1, k2) leading to an emission spectrum with three ZPLs i.e., V1 (864.7 nm), V2 (886.3 nm) and V3 (907.1 nm) \cite{sorman2000silicon, janzen2009silicon}. The h-V\textsubscript{Si}$^{^-}$ (V1 center) in 4H-SiC which has 8\% of emission into the ZPL at 4K \cite{udvarhelyi2020vibronic} and millisecond spin coherence times (under dynamical decoupling) \cite{nagy2018quantum, simin2017locking} provides avenues for high-fidelity spin-photon entanglement \cite{morioka2020spin} and deterministic spin initialization \cite{dong2019spin, nagy2019high}.  V\textsubscript{Si}$^{^-}$ is well-suited for photonic integration due to its single spin orientation along the c-axis for all inequivalent lattice sites \cite{sorman2000silicon, soltamov2015optically}, which is necessary for their deterministic orientation in devices. Moreover, the optical transitions in h-V\textsubscript{Si}$^{^-}$ are decoupled from charge fluctuations \cite{udvarhelyi2019spectrally} resulting in low spectral diffusion in an emitter ensemble \cite{nagy2021narrow} and making it robust to surface charge fluctuations in nanofabricated devices.  

The negatively charged nitrogen vacancy (N\textsubscript{C}V\textsubscript{Si}$^{^-}$) in SiC consists of a substitutional nitrogen atom at a carbon site adjacent to a silicon vacancy. The emission has multiple ZPLs corresponding to the inequivalent lattice sites with emission wavelengths close to the telecommunications range, making it suitable for low-loss fiber transmission \cite{zargaleh2016evidence, von2015identification, zargaleh2018nitrogen, khazen2019high}. A recent study has identified the emission wavelength of NV center in 3C-SiC at 1289 nm \cite{jurgen2021spin} correcting the previously reported value of 1468 nm \cite{zargaleh2018nitrogen}. Coherent control of NV spins at room temperature \cite{wang2020coherent}, robust photon purity, and photostability of single NV centers at elevated temperatures \cite{wang2020experimental} have been demonstrated. The possibility of coupling electron spin of NV-center to the nuclear spin of the substitutional nitrogen should be studied for quantum memory applications.

Other promising but less explored color centers in SiC are transition metal defects like chromium ion (Cr\textsuperscript{4+}) \cite{son1999photoluminescence, koehl2017resonant, diler2020coherent}, vanadium ion (V\textsuperscript{4+}) \cite{spindlberger2019optical, wolfowicz2020vanadium}, molybdenum ion (Mo\textsuperscript{5+}) \cite{gallstrom2009optical, bosma2018identification, gilardoni2020spin}, tungsten ion (W\textsuperscript{5+}) \cite{gallstrom2012optical} and erbium ion (Er\textsuperscript{3+}) \cite{tabassum2020engineered, parker2021infrared}. Some of these impurities have high DWF (Cr\textsuperscript{4+} - 73\%, V\textsuperscript{4+} - 22\% (h), 39\% (k)) suited for generating indistinguishable photons. Since the transition metals are not among the naturally occurring impurities in SiC, they are generated either by doping during the crystal growth or ion implantation on the wafer.

\subsection{Quantum optical properties of color centers}
Color centers create a set of ground and excited states within the bandgap of the host material and are often approximated as a two or a three level system. The optical properties of color centers are inspected in a photoluminescence or a photoluminescence excitation spectroscopy where the emitter is excited by an above-band or a resonant laser, respectively. The emission spectrum has two features: a sharp peak called the zero phonon line (ZPL) corresponding to a fully optical (no-phonon) transition, and a broad phonon sideband (PSB) involving transitions to the vibronic levels of the ground state. The ratio of light emitted into the ZPL is called the Debye-Waller Factor (DWF).

\begin{figure}[htbp]
    \centering
    \includegraphics[scale=0.5]{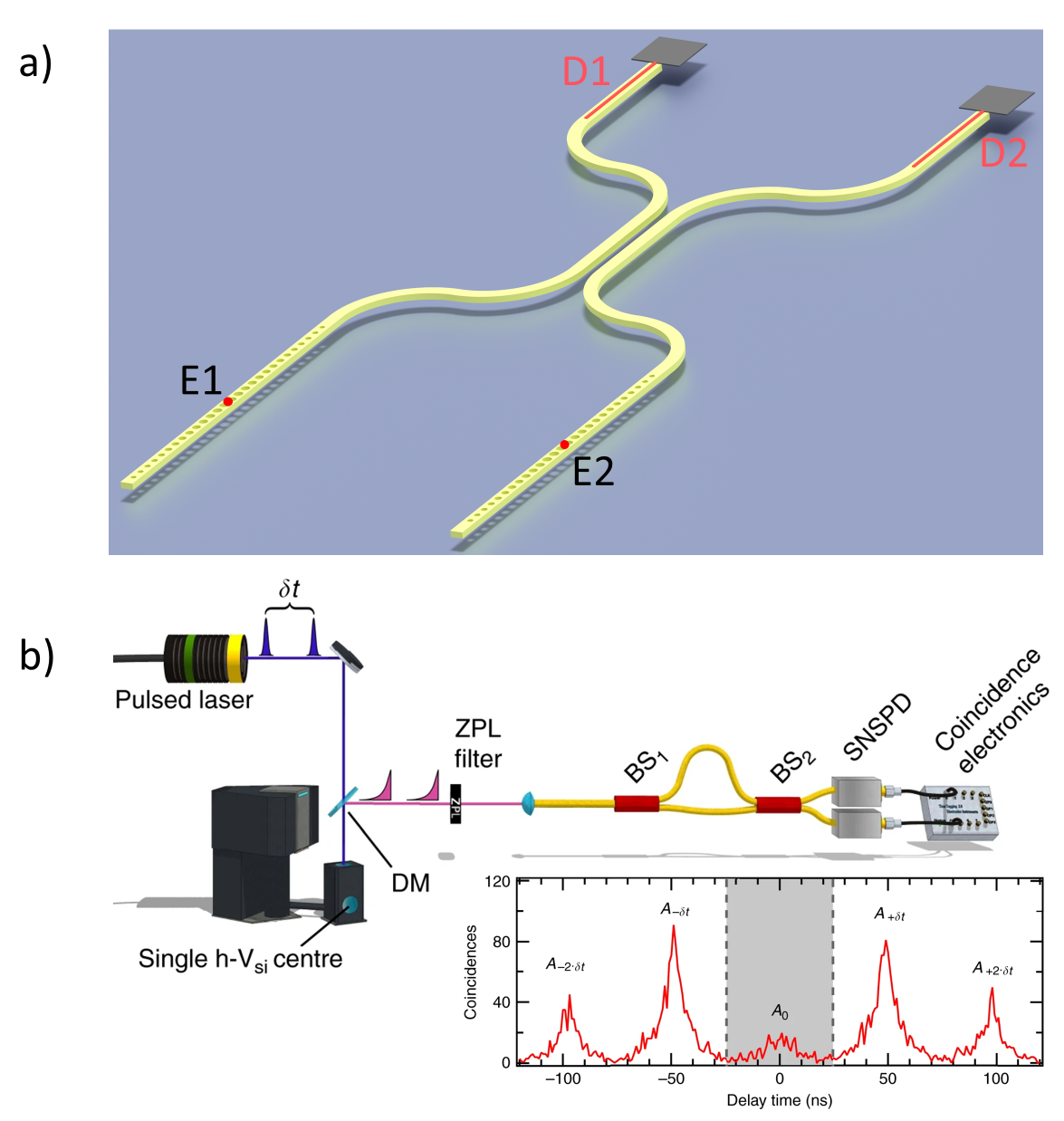}
    \caption{a) Perspective figure showing a concept of the on-chip realization of the HOM setup. Single color centers (E1, E2) are positioned at the center of an asymmetric cavity, resonant with the ZPL of the emitters. The emission into an asymmetric cavity preferentially propagates to one side coupling into the waveguide. The HOM interference occurs in the central region where the waveguide arms propagate in proximity mimicking a beam-splitter. The coincidence counts are measured using SNSPDs (D1, D2) that couple evanscently to the waveguide signal. b) (top) Schematic of a Hong-Ou Mandel interference setup to measure the indistinguishability of silicon vacancy in silicon carbide. (bottom) Plot showing two-photon coincidence counts as a function of time delay between the arrival of two photons at the superconducting nanowire single photon detectors (SNSPDs). It can be seen that the coincidence counts at zero delay has diminished, resulting in a visibility of 90\% \cite{morioka2020spin}. Reproduced with permission from Morioka, \textit{et al.}, Nat Commun 11, 2516 (2020); licensed under a Creative Commons Attribution (CC BY) license. 
    }
    \label{fig:fig2}
\end{figure}

Color centers are single photon emitters (SPEs) and as such suitable for applications in quantum light generation, spin-photon entanglement, and entanglement distribution. For these applications, it is desired to characterize the emitter's single-photon purity, which in ideal case yields a low probability of multi-photon emission and a high emission rate (brightness) \cite{aharonovich2016solid}. An SPE is a non-classical light source with sub-Poissonian statistics and anti-bunching behavior quantified by the zero-time second-order autocorrelation function ($g^{(2)}(\tau = 0))$, and higher order autocorrelation functions $g^{(n)}({\tau_i})$ \cite{stevens2014third, rundquist2014nonclassical}. The $g^{(2)}(\tau)$ function values are measured by the coincidence counts in a Hanbury-Brown-Twiss (HBT) interferometer. The HBT interferometer takes the signal from an optical mode and, using a 50:50 beam splitter, separates it into two paths terminated by single photon detectors with timed difference $\tau$ between detection events. As such, HBT is a reduced version of the Hong-Ou-Mandel interferometer (discussed in next paragraph) shown in Figure \ref{fig:fig2}a, where only one emitter is excited while the other one is left obsolete. If a color center is truly an SPE (and there is no luminescing background), the detectors will never have an immediate correlated detection event, yielding $g^{(2)}(\tau=0)=0$. In practice, however, $g^{(2)}(\tau=0)<0.5$ is used as a confirmation of the single-photon emission.

The coherent emission of color center light is possible through the ZPL, which presumably emits indistinguishable photons with identical spectral and temporal profiles. The Hong-Ou-Mandel (HOM) interferometer (Figure \ref{fig:fig2}b) is used to quantify this property with either subsequently emitted photons from the same emitter, or photons emitted simultaneously from separate emitters. When two indistinguishable photons are simultaneously incident on a 50:50 beam splitter, they undergo bunching. This results in a dip in the coincidence counts at the output detectors as shown in Figure \ref{fig:fig2}b. The degree of indistinguishability of the photons is quantified by the HOM visibility, determined by the depth of the dip in the coincidence counts. In a recent experiment with silicon vacancy in 4H-SiC \cite{morioka2020spin}, the measured visibility was $V=0.85$, which compares well to $V=0.90(6)$ obtained in diamond  for quantum networking applications \cite{hensen2015loophole}.

With advances in SiC material processing, an on-chip HOM interferometer, as shown in Figure \ref{fig:fig2}a, could be fabricated. The illustrated setup has three different regions, the cavities with color centers, the waveguide beam splitter, and the superconducting nanowire single photon detectors (SNSPDs). First, the ZPL emission from each of the color centers  (E1, E2) is enhanced in the cavity and coupled into a waveguide \cite{faraon2007efficient}. Second, the waveguides are laid out in close proximity to create coupling of optical modes. This region performs the function of a beam splitter. Third, the photons in each waveguide are detected using an SNSPD integrated on top of SiC \cite{martini2019single, martini2020electro}. SNSPDs offer advantages like high detection efficiency, low dark count rates, and low jitter over a broad wavelength range \cite{lita2008counting, marsili2013detecting, rath2015superconducting}. When two indistinguishable photons in each of the waveguides arrive simultaneously to the beam splitter region, they undergo bunching and are to be detected only at one of the two detectors (D1, D2). 

The coherence of a quantum system is susceptible to dephasing due to the thermal fluctuations in the environment, proximity to electron and nuclear spins, and population relaxation. The rate of dephasing of the state of a quantum system (spin) is determined by the coherence time, during which the quantum system can be reliably manipulated and measured. Systems with long coherence times would allow for the application of more quantum operations, required to build large scale quantum systems with hundreds of qubits \cite{wolfowicz2021quantum}. For typical spin-based two-qubit quantum gates which have a predicted gate operation time of $\sim500$ ns \cite{dong2020precise}, to perform 100,000 gate operations on the quantum state, needed for the implementation of quantum error correction \cite{divincenzo2000physical}, a coherence time of the order of 50 ms is required. The effect of dephasing mechanisms can be minimized using techniques like isotopic purification and dynamical decoupling. Another option to extend coherence times is using gate operations to transfer the electron spin onto a nuclear spin memory \cite{maurer2012room, bradley2019ten}.

In addition to the long spin coherence that allows for complex manipulations, the quantum gates should also reach high fidelities. Recent demonstrations in divacancy centers \cite{bourassa2020entanglement} achieve 99.984\% single-qubit gate fidelity, entangling CNOT gate fidelity of 99\% and entangled state generation fidelity of 81\%. In a recent silicon vacancy study \cite{babin2021fabrication}, the projected values for the single-qubit X gate and the Bell state preparation fidelities are 97\% and 94\% respectively. For comparison, Bell state violation experiments in diamond used systems with fidelities for Bell state preparation of 92\% and X gate of 99.8\%. The quantum repeater proposals \cite{jiang2009quantum} target fidelity of 95\% for Bell state creation. This shows that silicon carbide divacancy and silicon vacancy centers are approaching the performance suitable for quantum networks.

\subsection{Motivation for QIP implementations as a SiC platform}
After the earliest demonstration of NV center in diamond as a quantum system that can be individually initialized, manipulated and measured with high fidelity at room temperature \cite{neumann2008multipartite}, this spin-1 emitter has been the candidate for majority of the important demonstrations in the field of QIP. Despite being the traditionally most widely studied color center, diamond NV has several shortcomings including a low DWF of 3\% \cite{faraon2011resonant, riedel2017deterministic} that prevents achieving high entanglement rates, unavailability of single-crystal diamond wafers, emission wavelength far outside the telecommunication wavelength range (637 nm), and large spectral shifts caused by fluctuations of charges in its surrounding \cite{doherty2013nitrogen}. The spectral shifts are further pronounced by proximity to surfaces \cite{ruf2019optically}, which makes this emitter challenging for efficient nanophotonic integration. It is worth noting that quantum frequency conversion of NV center emission to telecommunication wavelengths with preserved spin-photon entanglement is possible \cite{tchebotareva2019entanglement}. Other color centers explored in diamond include group-IV color centers like silicon vacancy \cite{sukachev2017silicon, pingault2017coherent}, germanium vacancy \cite{siyushev2017optical}, tin vacancy \cite{iwasaki2017tin, trusheim2020transform} and lead vacancy \cite{trusheim2019lead}. These exhibit higher DWF and better optical stability, however, their emission wavelengths are far from the telecommunication bands and they have not yet demonstrated long coherence times except at mK temperatures \cite{sukachev2017silicon, siyushev2017optical}. 

Silicon carbide hosts a variety of color centers \cite{atature2018material, son2020developing} with long spin coherence times \cite{widmann2015coherent, christle2015isolated, seo2016quantum}, excellent brightness \cite{castelletto2014silicon} and manipulations of nuclear spin \cite{falk2015optical, klimov2015quantum}. SiC offers benefits like large bandgap, high thermal conductivity, and strong second-order nonlinearity. It also hosts color centers at infra-red wavelengths and is available in several polytypes (3C-SiC, 4H-SiC, 6H-SiC) offering flexibility in terms of material properties and color center configurations. Due to its decades-long industry presence, SiC is available as high-quality single crystal wafers, has mature nanofabrication processes, and is CMOS process compatible. Most importantly, SiC color centers have longer spin coherence times than the ones in the naturally occurring diamond \cite{seo2016quantum, yang2014electron} and on par with isotopically purified diamond.

\section{Integration with photonics}
Color center properties are sensitive to its environment (neighboring crystal defects, nuclear spins, surface related charges, and coupling to phonons) which poses a challenge to their naturally available forms in achieving theoretically calculated spin-optical properties. In this section, we will look into different approaches to mitigate the influence of environment on the color center properties through material engineering and integration with nanophotonics, to achieve improved spin-optical characteristics (more indistinguishable photons, efficient optical initialization of spin, longer coherence times, on-chip integration and scalability) and better spin-photon interfaces for applications in QIP.

\subsection{Isotopic Purification}
Naturally available SiC has 1.1\% \textsuperscript{13}C and 4.7\% \textsuperscript{29}Si atoms, both with a nuclear spin of \textit{I} = 1/2. The electron spin of the color centers and the nuclear spins have an always-on hyperfine interaction. This interaction can be used to transfer and store the electron spin state in a nuclear spin \cite{yang2016high} (quantum memory), which has a long coherence time, on the order of a few seconds, necessary for QIP applications like error correction and entanglement purification. But, the electron spin immersed in the nuclear spin bath experiences decoherence due to fluctuating nuclear spins, that results in a nuclear spin decoherence of the nearby nuclear spin. This issue can be addressed using isotopic purification, which is the careful engineering of the nuclear environment of the color centers. A detailed study of the optimal nuclear spin concentration for maximum usable nuclear memories was done \cite{bourassa2020entanglement}. Isotopically purified SiC \cite{simin2016all} was used to demonstrate millisecond electron spin coherence times \cite{seo2016quantum, morioka2020spin, bourassa2020entanglement, nagy2019high}, coupling to single nuclear spin with 1 kHz resolution \cite{nagy2019high}, and high-fidelity electron spin control (99.984\%) \cite{bourassa2020entanglement}.

\subsection{Passive and active photonic devices}
To build a large quantum network with photon-based entanglement distribution, an important criterion is the availability of network nodes with efficient spin-photon interfaces. This can be achieved by efficiently collecting the photons emitted by a color center in the bulk and by enhancing the proportion of photons emitted into the ZPL. Passive devices couple light into an optical mode to improve collection efficiencies and low-loss guiding of the photons. Active devices improve the spontaneous emission rates of the color centers.  

Color center emission in a high refractive index bulk is directed based on its dipole orientation, and when a fraction of this emission reaches the bulk-air interface, it undergoes total internal reflection, resulting in low collection efficiencies of the emitted photons. The most implemented approach for improving collection efficiencies is fabricating a solid immersion lens (SIL) on top of the color center, this ensures that the color center emission has a normal incidence at the bulk-air interface. SIL fabricated in SiC on top of a silicon vacancy \cite{widmann2015coherent, sardi2020scalable} demonstrated collection enhancements up to 3.4 times. Similar enhancements were achieved using scalable nanopillars with integrated silicon vacancies \cite{radulaski2017scalable, lukin20204h} and erbium color centers \cite{parker2021infrared}. The collection efficiencies can be further enhanced by deterministic emitter positioning within these devices. Nanofabricated waveguides can be used for coupling color center emission into well-defined modes, to be routed on a chip in a nearly lossless way. High-coherence spin-optical properties of silicon vacancy integrated into SiC waveguides was demonstrated recently \cite{babin2021fabrication}. The emission in the waveguide mode can be coupled into a fiber for off-chip processing using free space grating couplers \cite{dory2019inverse} and tapered fiber couplers \cite{burek2017fiber} that were fabricated in diamond and yet to be developed for SiC.

\begin{figure}[ht!]
    \centering
    \includegraphics[scale=0.5]{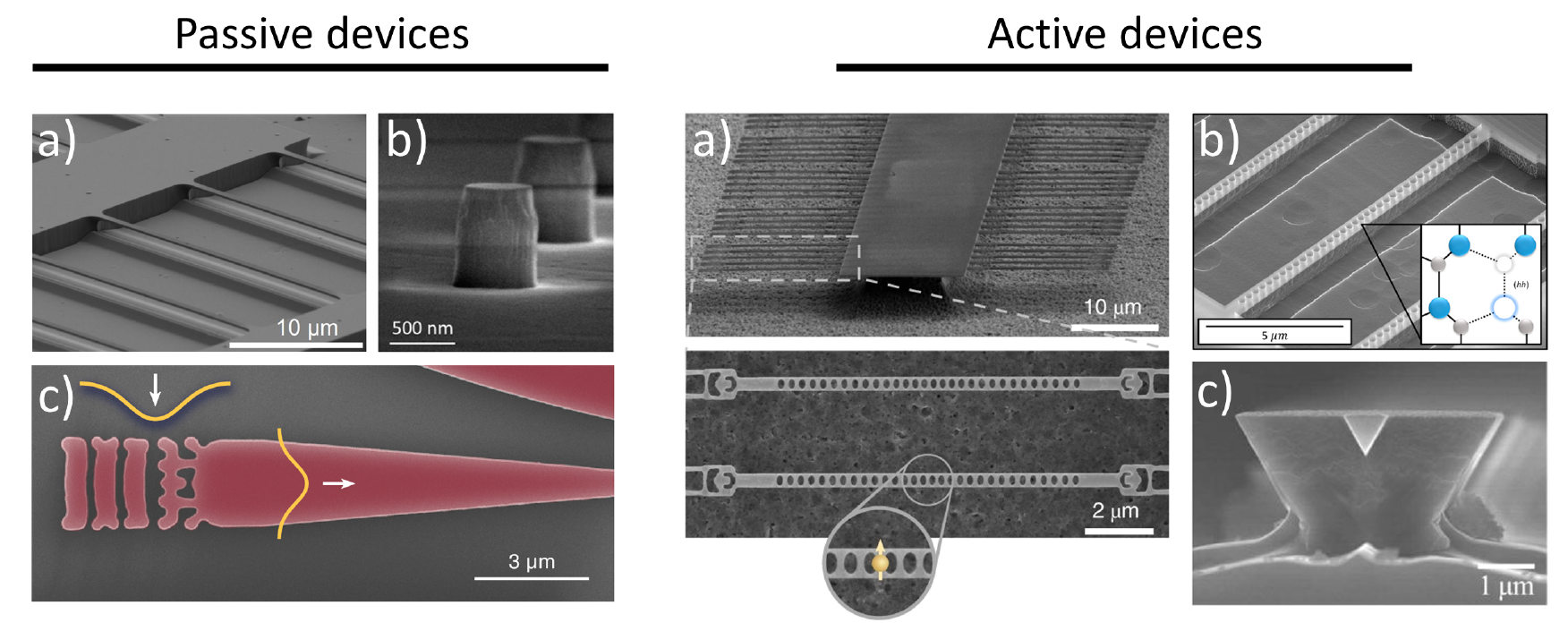}
    \caption{The SEM images of silicon carbide photonic devices. \textbf{Passive devices} (clockwise from top left): a) Triangular cross-section waveguide in 4H-SiC with integrated silicon vacancies (Reproduced with permission from Nat. Mater. 21, 67–73 (2022). Copyright 2021, Springer Nature.), b) nanopillars in 4H-SiC with silicon vacancies (Reproduced with permission from Nano Lett. 17(3), 1782–1786 (2017). Copyright 2017, American Chemical Society.), c) vertical coupler fabricated in 4H-SiCOI to convert free-space Gaussian beam into fundamental waveguide mode (Reproduced with permission from Optica 7, 1139-1142 (2020).); \textbf{Active Devices:} a) Suspended nanobeam array fabricated in 4H-SiCOI and (below) magnified image of individual nanobeam PCC with integrated silicon vacancies (Reproduced with permission from Nat. Photonics 14, 330–334 (2020). Copyright 2019, Springer Nature.), b) nanobeam PCC in 4H-SiC with integrated neutral divacancies (Reproduced with permission from Nano Lett. 20 (5), 3427–3434 (2020). Copyright 2020, American Chemical Society.), c) triangular cross-section nanobeam PCC fabricated in 4H-SiC \cite{song2018high} (Reproduced from Appl. Phys. Lett. 113, 231106 (2018), with the permission of AIP Publishing.).   
    }
    \label{fig:fig3}
\end{figure}

Active devices like photonic crystal cavities (PCC) create light-matter interaction by confining light to sub-wavelength volumes, resulting in an increased density of states at the resonant wavelength of the cavity. When the ZPL wavelength of the color center is resonant with the cavity, the spontaneous emission rate of the color center is enhanced through Purcell effect. The enhancement in radiative emission is given by Purcell factor (\textit{F\textsubscript{P}}):
\[
    F_P = \left[\frac{3}{4\pi^2} \bigg(\frac{\lambda}{n} \bigg)^{3} \bigg(\frac{Q}{V} \bigg) \right]\left[ \left| \frac{E}{E_{\textnormal{max}}} \right| \cos(\phi) \right] \xi
\]
\textit{F\textsubscript{P}} can be maximized by integrating a color center into a cavity with high quality factor (\textit{Q}) and low cavity mode volume (\textit{V}). \textit{F\textsubscript{P}} can be further boosted by optimally positioning the emitter inside the cavity, which corresponds to the spatial overlap ($\left| E/E_{\textnormal{max}} \right| \cos(\phi)$) term in the equation. It should be noted from the equation above that the \textit{Q}/\textit{V} value required to achieve the same \textit{F\textsubscript{P}} value for different color centers depends on the DWF of the color center ($\xi$). Several cavity geometries like 1D nanobeam PCC \cite{bracher2017spontaneous, crook2020purcell, lukin20204h}, 2D PCC \cite{song2011demonstration, radulaski2013photonic, song2019ultrahigh}, microdisk resonator \cite{lu2013silicon, magyar2014high}, triangular cross-section nanobeams \cite{song2018high} were fabricated in SiC, as shown in Figure \ref{fig:fig3} (right). V1 silicon vacancy \cite{lukin20204h, bracher2017spontaneous}, neutral divacancy \cite{crook2020purcell}, and Ky5 color centers \cite{calusine2014silicon, calusine2016cavity} coupled to SiC cavities demonstrated Purcell factors of up to $F_P=120$. 

Purcell factor considers only the radiative emission resonant with the cavity, which might not be the only path for the emitter to decay from the excited state. Other pathways like non-radiative decay, emission into the phonon sideband, and pure dephasing might exist. Cooperativity (\textit{C}) is the ratio of radiative emission rate through the cavity to the total decay rate of the emitter in the absence of the cavity \cite{janitz2020cavity, borregaard2019quantum}. Although the spontaneous emission is enhanced through Purcell effect (\textit{F\textsubscript{P}} $>$ 1), the emission through the cavity can still be smaller compared to the overall emission - weak coupling regime (C $<$ 1). To achieve deterministic atom-photon interactions, the coupling between the cavity and emitter should be stronger than all the other dephasing mechanisms - strong coupling regime (\textit{C} $>$ 1) \cite{reiserer2015cavity, borregaard2019quantum}. Further, emitter ensembles in a cavity with collective coupling rates to the cavity resonant mode comparable to the inhomogeneous broadening exhibit longer storage times, apt for quantum memory applications \cite{diniz2011strongly}.  

Putting these concepts into the context of QIP, to reach 99\% coherent emission (through the ZPL) of a color center with DWF of 5\% and fully optical recombination (no non-radiative recombination channels), the emission would need to be  Purcell enhanced by a factor $F_P\sim2,000$. To achieve this in a representative nanocavity with a cubic wavelength mode volume $V=(\lambda / n)^3$ would require a quality factor of $Q\sim5\times 10^5$. Assuming now that there is a non-radiative decay channel with three times the rate of the radiative recombination, the nanocavity would need to achieve $F_P\sim8,000$ and $Q\sim 2\times 10^6$. Related calculations for color centers in SiC are reported in Table \ref{tab:Table1}. It should be noted that for color centers with closely spaced ZPLs, such as V2 silicon vacancy, additional linewidth broadening and spectral overlap should be considered when evaluating coherence properties of Purcell enhanced emission. Q-factors up to $6.3\times10^5$ \cite{song2019ultrahigh} have already been achieved in SiC nanocavities without emitters. Thus, there needs to be more research focus on color center integration into the SiC nanocavities. 

\begin{table}[htbp]
\centering
\resizebox{\columnwidth}{!}{\begin{tabular}{*9c}
\hline

Defect & ZPL & DWF & $T_{2}$ & Fiber loss &\multicolumn{2}{c}{Propagation length in fiber (km)} & \multicolumn{2}{c}{Target Q-factor for $V=(\lambda/n)^3$}\\ \cmidrule(lr){6-7} \cmidrule(lr){8-9}
&(nm)&& (ms) & (dB/km) & before 1/\textit{e} loss & during $T_2$ & radiative decay only & 3:1 non-rad.:rad. decay\\
\hline
NV$^{^-}$(diamond) & 637 & 3\% & 1.8 & 10 & 0.4 & 374 & $1.4\times10^6$ & $5.8\times10^6$\\
V\textsubscript{Si}$^{^-}$(4H-SiC) & 861-917 & 6-9\% & 20 & 1.5 & 2.9 & 4,160 & $(0.15-0.34)\times10^6$ & $(0.6-1.4)\times10^6$\\
V\textsubscript{Si}V\textsubscript{C}$^0$(4H-SiC) & 1078-1132 & 5\% & 5300 & 0.8 & 5.4 & $1.1 \times 10^6$ & $0.5\times10^6$ & $2.1\times10^6$\\
N\textsubscript{C}V\textsubscript{Si}$^{^-}$(4H-SiC) & 1176-1242 & 5\% (3C) & 0.017 & 0.6 & 7.2 & 3.53 & $0.5\times10^6$ & $2.1\times10^6$\\

\hline
\end{tabular}}
\caption{Emission parameters for NV center in diamond, and silicon vacancy (V\textsubscript{Si}$^{^-}$), divacancy (V\textsubscript{Si}V\textsubscript{C}$^0$), and nitrogen vacancy (N\textsubscript{C}V\textsubscript{Si}$^{^-}$) in 4H-SiC. $T_2$ is the electron spin-coherence time. The highest measured electron spin-coherence times\cite{balasubramanian2009ultralong, simin2017locking, anderson2021five, wang2020coherent} and DWF\cite{riedel2017deterministic, udvarhelyi2020vibronic, crook2020purcell, jurgen2021spin} are presented. 1/$e$ loss corresponds to the power reduction by the factor $e=2.718$. The distance covered by the photon during $T_2$ is calculated by assuming a constant speed of light in the fiber (refractive index = 1.44) at all wavelengths. The target Q-factors to achieve 99\% emission through the ZPL are calculated for a cavity with mode volume of $(\lambda/n)^3$ for cases of purely radiative recombination and 3:1 non-radiative to radiative decay ratio.}
\label{tab:Table1}
    
\end{table}

\subsection{Nanofabrication Methods}
To build scalable QIP technologies, photonic devices need to be fabricated in a repeatable manner on chip-scale. Despite the mature nanofabrication processes in SiC, the substrate is chemically inert and difficult to etch through wet-etching methods. Dry etching methods \cite{jiang2004impact, choi2012fabrication, dowling2016profile} used to fabricate photonic devices in SiC were not able to achieve simulated ultra-high values of Q-factor due to fabrication errors (misalignment of features), surface roughness (low Q-factor), and imperfect emitter positioning in the cavity (poor coupling between cavity and emitter). Surface passivation using epitaxial AlN was used to improve the photostability, charge state stability, and enhance the emission of carbon anti-site vacancy in 4H-SiC \cite{polking2018improving}.

Integration of color center photonics with other platforms starts with heteroepitaxially grown high quality single crystal thin films on silicon and silicon dioxide. The low temperature phase, 3C-SiC, which readily grows on silicon substrates was used to fabricate photonic devices \cite{cardenas2013high, radulaski2013photonic, lu2014high}. But, these 3C-SiC thin films had high densities of crystal defects due to the lattice mismatch with Si, resulting in large material absorption and degradation in color center optical properties \cite{calusine2016cavity}. Smart-cut method used to create SiC-on-insulator (SiCOI) \cite{di1997silicon} is not suitable for color center photonics as the SiC layer suffered from implantation damages that are irreparable through annealing at elevated temperatures. Recently, 4H-SiCOI was produced by wafer bonding 4H-SiC on insulator (silicon dioxide) followed by grinding and polishing the SiC to obtain high-quality thin films suitable for fabricating high performing photonic devices and color center integration \cite{lukin20204h}. While successful in generating high-quality individual devices, this process faces challenges in achieving uniform film thickness through polishing on a wafer-scale.

Underetching of photonic structures which creates optical isolation and prevents mode leakage into the bulk was used to demonstrate PCC with high Q-factor ($> 10^{4}$). Dopant selective photoelectrochemical etch \cite{bracher2015fabrication} was used to fabricate 1D nanobeam PCC with Purcell enhancement up to 80 times \cite{bracher2017spontaneous, crook2020purcell}. But, this method is limited due to the requirement of specific doping profiles in SiC. Angled etching is another underetching method first demonstrated in diamond \cite{burek2012free} and later demonstrated in SiC \cite{song2018high, babin2021fabrication}. The triangular cross-sections resulting from the angled etching in SiC was achieved using Faraday cage method and can be done at wafer scale using ion beam etching methods used in diamond \cite{atikian2017freestanding}.

Nanofabrication process variability results in deviation of the cavity's resonant wavelength designed to match the ZPL wavelength of the color center. The cavity can be tuned into resonance by condensing gas on the devices \cite{mosor2005scanning, zhang2018strongly} and effectively reversed by heating the chip. For a much more precise and fast tuning, the emitter can be tuned into resonance with the cavity using DC Stark effect \cite{anderson2019electrical, ruhl2019stark, de2017stark, bathen2019electrical} and strain fields \cite{meesala2018strain, machielse2019quantum}. Both these methods can be used to tune spatially separated emitters into mutual resonance necessary for remote entanglement. Despite advanced fabrication processes and high crystal quality, the emitters experience spectral diffusion (broadening of optical linewidths) due to charge fluctuations in the local environment caused by surface charges, and distribution of emitters which capture and release charges. Strain generated during growth and nanofabrication of photonic devices also contributes to the degradation of optical linewidths. But this spectral diffusion/instability is an extrinsic property that can be solved either by integrating charge depletion using advanced doping epitaxy or by optimized optical excitation via time-dependent drive or optimized steady-state illumination \cite{lukin2020integrated}. The charge state of the emitters were successfully controlled through optical illumination \cite{wolfowicz2017optical} and integration of emitters into Schottky barrier diodes \cite{bathen2019electrical}. Emitters integrated into the depletion region of a p-i-n diode enabled deterministic control over the charge state of the emitter \cite{widmann2019electrical, anderson2019electrical} and electrical readout of the spin state \cite{niethammer2019coherent}. Optimal emitter distribution can be achieved using more controllable defect generation methods like focused ion beam \cite{wang2017scalable}, laser writing \cite{chen2019laser, almutairi2022direct}, proton beam writing \cite{kraus2017three}, and ion implantation on patterned substrates \cite{wang2017efficient}.        

\subsection{External fields}
Upon addressing the inhomogeneities in the color center environment through isotopic purification and integration into electronic and photonic devices, the single photon emission of a color center can be tuned by a direct application of an external interaction such as electric field, magnetic field, mechanical strain, and acoustic fields \cite{wolfowicz2021quantum, bathen2021manipulating}. The energy levels of the color center are sensitive to the applied external field, causing a split or shift of the ground and/or excited states (tuning). For QIP applications, the tuning achieved by the application of electric or strain fields should be larger than the inhomogeneous broadening ($\sim10$ GHz for V\textsubscript{Si}$^{^-}$ \cite{babin2021fabrication}), before the material experiences any dielectric or mechanical failure \cite{wolfowicz2021quantum}.

The splitting and shifting of energy levels due to an electric and magnetic field are called the Stark and Zeeman effect. Electric fields have been used to tune the ZPL of silicon vacancy \cite{bathen2019electrical, ruhl2019stark} and divacancy \cite{anderson2019electrical, de2017stark} in 4H-SiC by up to 60 GHz and 800 GHz, respectively. In another take, a magnetic field applied along the c-axis of 4H-SiC was used to suppress the parasitic spin mixing of silicon vacancy, to achieve controlled generation of indistinguishable photons \cite{morioka2020spin}. Strain that is applied externally or formed intrinsically during crystal growth can result in tuning of the emission wavelength of color centers. Moreover, the density-functional theory predicts that silicon vacancy \cite{udvarhelyi2020vibronic} and divacancy \cite{falk2014electrically, udvarhelyi2018ab} in SiC produce larger ZPL shifts under strain fields than electric fields \cite{bathen2019electrical, ruhl2019stark, anderson2019electrical, de2017stark}. Silicon vacancy in 6H-SiC has been shown to exhibit strain-induced ZPL shifts up to 6 THz \cite{vasquez2020strain, breev2021stress}. In an engineered acoustic device, the coupling of spin to strain was demonstrated for divacancy in SiC \cite{whiteley2019spin}. This shows that the engineering toolbox has multiple methods to counter the inhomogeneous broadening of color center emission at or beyond the required magnitudes.

\section{Quantum Information Hardware Applications}
In this section, we discuss methods of employing color centers for applications in quantum networking, simulation, and computing.

\subsection{Quantum Repeaters}
In classical communication, the loss experienced by the signal (photons) propagating in the fiber is compensated through its amplification at intermediate points in the channel, thus enabling long-distance communication. In the quantum domain, communication over long distances cannot be implemented in such a fashion due to the consequences of the no-cloning theorem. In addition to the fiber losses (which scale exponentially with distance between entangled links), the photons in the quantum communication channel also experience depolarization losses. A long range entanglement can be generated using quantum repeater (QR) schemes which involve dividing the communication channel into smaller segments connected by the QR stations, followed by establishing entanglement between the ends of each of the segments. Then, the desired long range entanglement is generated by performing entanglement swapping of the entangled states of each segment. 

\begin{figure}[ht!]
    \centering
    \includegraphics[scale=0.7]{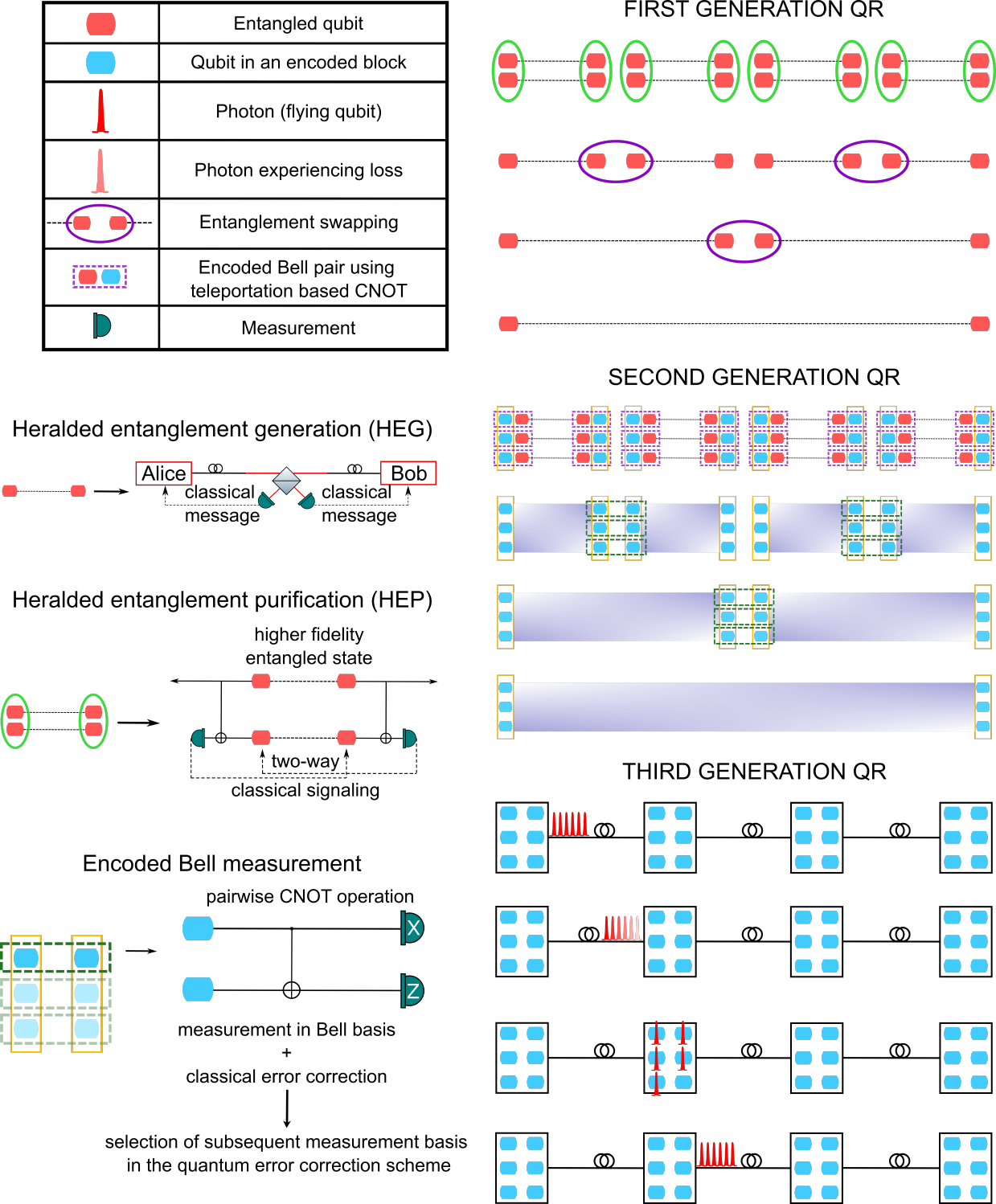}
    \caption[Three generations of quantum repeaters.] {The entanglement schemes designed in the three generations of quantum repeaters as a guide for SiC quantum information processing \cite{pompili2021realization, bhaskar2020experimental, hensen2015loophole, jiang2009quantum, munro2010quantum, abobeih2021fault, fowler2010surface, munro2012quantum, muralidharan2014ultrafast}.
    }
    \label{fig:fig4}
\end{figure}

The QR stations are designed to perform some (or all) of the following three tasks: heralded entanglement generation (HEG) for correcting loss errors, heralded entanglement purification (HEP) to correct for operation errors (from measurements and gate operations), and entanglement swapping to extend the entanglement. Entanglement (HEG) is generated between color centers in two remote QR stations by first creating spin-photon entanglement locally, followed by interference of the photons on a beam splitter to create spin-spin entanglement \cite{klimov2015quantum, bourassa2020entanglement, nguyen2019integrated}, as shown in Figure \ref{fig:fig4}. HEP using color centers begins with swapping a low-fidelity electron spin-spin entanglement onto nuclear spins (memory qubits), followed by another cycle of HEG (low-fidelity), and finally local gate operations are performed on the electron and nuclear spin in each QR station to generate a higher fidelity entanglement link \cite{kalb2017entanglement, lu2020heralded}. 

The heralding involves two-way signaling between QR stations to verify the success of entanglement/purification, which requires long-lived quantum memories \cite{munro2015inside} and determines the achievable communication rates. Instead, quantum error correction (QEC) can be used to overcome the loss and operation errors deterministically, requiring one-way signaling. Depending on the approach for overcoming losses, quantum repeaters can be classified into three generations, each offering improved communication rates over implementations without quantum repeaters, which follows an inverse exponential scaling with the total distance between the ends of the network.

The first generation QRs overcomes loss and operation errors through HEG and HEP respectively. The communication rate (1 Hz range) is limited by the two-way signaling and would require quantum memories with milliseconds or higher coherence times for longer distances (> 1000 km), even for optimized protocols \cite{jiang2007optimal}. The second generation QRs, like the first one, overcomes loss through HEG but uses QEC on encoded qubits\cite{jiang2009quantum, munro2010quantum, abobeih2021fault} to rectify operation errors. The communication rate (kHz range) depends on the time needed for two-way signaling between adjacent QR stations and local gate operations, relaxing the long coherence time requirement for memory qubits. The third generation QRs uses QEC for overcoming both loss and operation errors \cite{fowler2010surface, munro2012quantum, muralidharan2014ultrafast}. Since two-way signaling is not involved in this generation, the communication rates depend on the local gate operation times, achieving rates (MHz) close to the classical repeaters. However, this generation suffers from low tolerance to matter-photon coupling losses ($\lesssim$ 10\%) even with hundreds of qubits per repeater station \cite{muralidharan2014ultrafast}.

The successful implementation of the different QR generations depends on the various parameters of the technological building blocks in their architecture. This includes the fidelity and operation time of the gates performing local operations, as well as the coupling efficiency (emission of photons from matter qubits, coupling of photons into the fiber, coupling between photons and matter qubits, detection efficiency). Cost function \cite{muralidharan2014ultrafast} analysis shows that a) high gate error probability ($\gtrsim 1\%$) is more suited for the first generation QRs, b) low/intermediate gate error probability ($\lesssim 1\%$), low coupling efficiency ($\lesssim 90\%$), and slow gate operation ($\gtrsim 1 \mu$s) benefit the second generation QRs, and c) low gate error probability ($\lesssim 1\%$), high coupling efficiency ($\gtrsim 90\%$), and fast gate operation ($\lesssim 1 \mu$s) favors the third generation QRs \cite{muralidharan2016optimal}.

Color center systems in SiC used to demonstrate HEG \cite{klimov2015quantum, bourassa2020entanglement} are suitable for implementation of the first and the second generation of QRs. Their potential for long-distance emission propagation and the technological requirements for highly indistinguishable photon generation are  analyzed in Table \ref{tab:Table1}. Spectral inhomogeneity of disparate emitters can be improved by 1) minimizing spectral diffusion through optical illumination \cite{wolfowicz2017optical} and integration of color centers into Schottky \cite{bathen2019electrical}, p-i-n \cite{widmann2019electrical, anderson2019electrical} diodes, and 2) Emission tuning through the application of electric field \cite{bathen2019electrical, ruhl2019stark, anderson2019electrical, de2017stark} and strain \cite{udvarhelyi2020vibronic, falk2014electrically, udvarhelyi2018ab, vasquez2020strain, breev2021stress}. Isotopic purification and dynamical decoupling should be employed to assure long spin-coherence times. Color centers with higher spin-coherence times (seconds range) can be used to implement the first generation QRs for longer distances ($\gtrsim 1000$ km). The emission rates can be improved by efficient integration of color centers into nanophotonic devices through Purcell effect. As discussed in the previous section, nanophotonic integration can be further improved through triangular cross-section devices, optimal positioning of color center in the cavity, and improved nanofabrication to reduce scattering losses. Efficient nanophotonic integration also boosts the spin-photon coupling efficiency necessary for implementation of the third generation QRs. To complete the tool set, the HEP obtained with NV center in diamond \cite{kalb2017entanglement} is yet to be demonstrated using SiC color centers.

Critical steps in the implementation of entanglement protocols like high-fidelity spin state initialization and resonant readout \cite{bourassa2020entanglement, christle2017isolated}, single-shot readout via spin-to-charge conversion \cite{anderson2021five}, and high-fidelity nuclear spin initialization \cite{bourassa2020entanglement, ivady2016high} were demonstrated using neutral divacancies in SiC. Despite the low DWF of divacancy ($\sim$5\%), single-shot readout measurements can be pursued through efficient cavity integration \cite{crook2020purcell}. Silicon vacancy has a strong intersystem crossing which allows for high-fidelity spin initialization \cite{nagy2019high}, but also prevents high-fidelity readout \cite{banks2019resonant}. Here too,  integration into a nanocavity with moderate Purcell factor of less than 100, can result in photon counts sufficient to achieve quantum non-demolition readout of the spin state \cite{nagy2019high}. Efficient quantum frequency conversion of emission from divacancy and silicon vacancy to the telecommunications band should be developed to extend the range of quantum communication. Integration of SNSPDs onto photonic devices would boost the success rates of the measurement protocols \cite{martini2019single, martini2020electro}. Other potential color centers that need to be studied closely are the ones with emission in telecommunications band (NV$^{-}$, V\textsuperscript{4+}, Er\textsuperscript{3+}) to minimize the fiber losses, and those with large DWF (Cr\textsuperscript{4+}, V\textsuperscript{4+}) that offer high photon counts for single-shot readout. It should be noted that the current state-of-the-art performance of color centers in SiC (especially divacancy) are more promising than the NV center in diamond for QIP applications. SiC photonic devices require lower Q-factors for achieving high photon counts and their color center's ZPL photons propagate for much longer distances in fiber, as shown in Table \ref{tab:Table1}.

The demanding requirements on the matter qubits for the implementation of quantum repeaters presents an opportunity for an all-photonic quantum repeater scheme implemented using single-photon sources, linear optical elements, photon detectors, and optical switches \cite{azuma2015all}. Realization of the highly entangled states known as graph states (Figure \ref{fig:fig6}a) are needed to implement an all-photonic measurement-based quantum repeaters \cite{zwerger2012measurement, zwerger2013universal}. A photonic cluster state can be generated by continuous optical pumping of a single photon emitter to produce a chain of photons that are entangled with other photons and the emitter \cite{buterakos2017deterministic}. Assuming millisecond spin coherence and nanosecond optical lifetime, a back of the envelope calculation would predict cluster states of the order of $10^6$ photons. However, to make this estimate more realistic to experimental conditions, the effects of non-radiative recombination and photon collection would need to be considered. These reductions could be made up for by Purcell enhancement in nanocavities. Approaches for generating large clusters of entangled photons for divacancy, silicon vacancy, and NV center in SiC have been proposed \cite{sophia2016spin}. To deterministically generate cluster states that outperform matter qubit based QRs, very high fidelity and fast entangling gates, and high photon generation and collection rates will be required \cite{hilaire2021resource}. 

\subsection{Quantum Simulation}
\begin{figure}[ht!]
    \centering
    \includegraphics[scale=0.5]{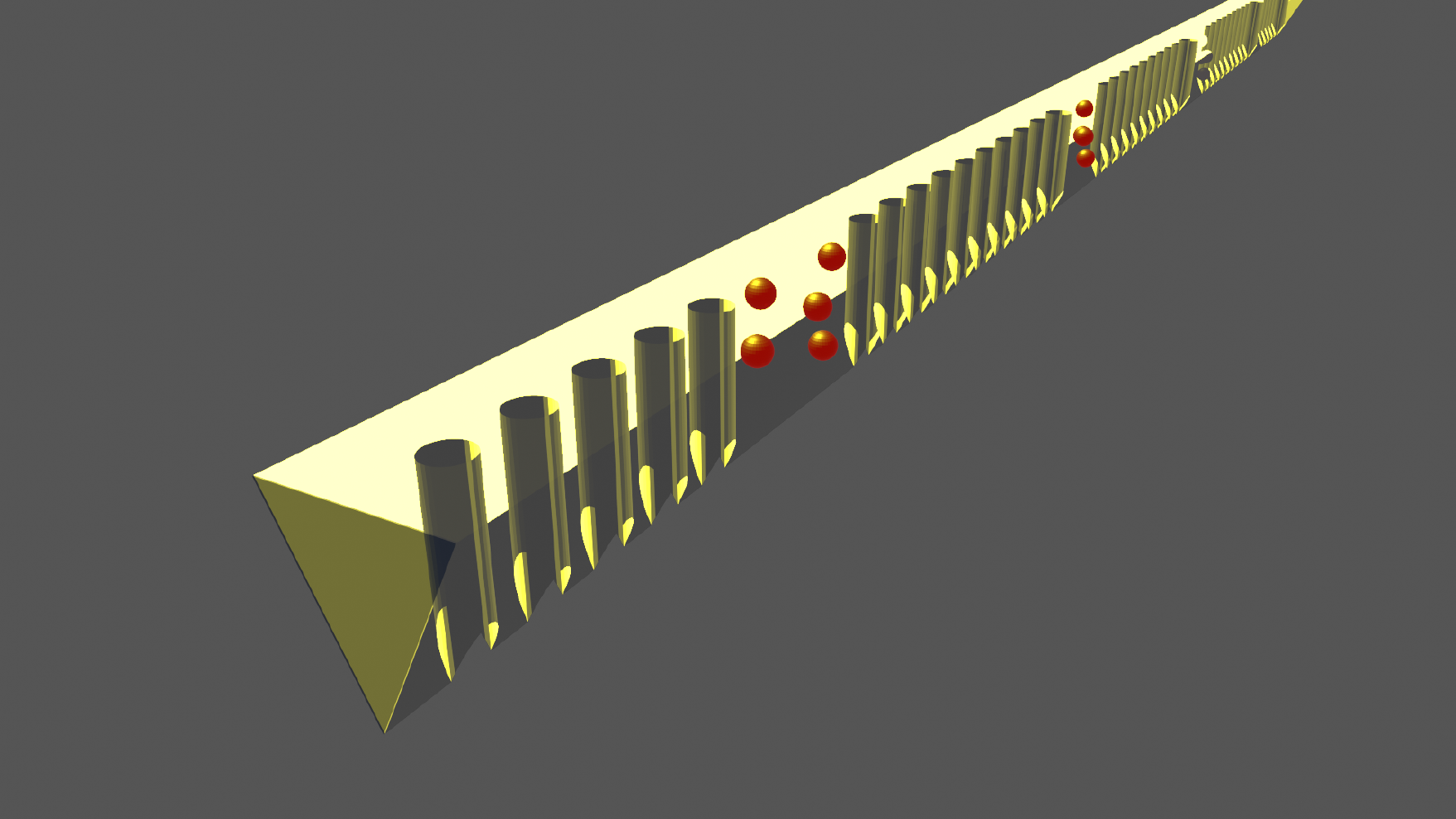}
    \caption{An illustration of an all-photonic quantum simulator that could be implemented in a triangular SiC coupled cavity array with integrated color centers \cite{majety2021quantum, patton2021all}.}
    \label{fig:TCHl}
\end{figure}

The idea of quantum information systems as models for quantum materials simulation has been around since the first quantum computing proposals. While a quantum advantage in simulation has not yet been demonstrated, with advancements in emitter integration into nanocavities, these nanophotonic systems are progressing toward versatile cavity quantum electrodynamical (CQED) platforms. Coupled cavity arrays with integrated emitters, as illustrated in Figure \ref{fig:TCHl}, have been of special interest for all-photonic quantum simulation. The modeling of such systems has shown their representation potential for the dynamics of complex condensed matter phases such as charge density waves, supersolidity, and the quantum phase transitions between the superfluid and Mott insulator phases \cite{hartmann2008quantum,bujnowski2014supersolid}. CQED-based quantum information frameworks primarily rely on the strong light and matter interaction, where the photon loss rates from the cavity and emitter, $\kappa$ and $ \gamma$ respectively, are less than twice the emitter-cavity coupling strength, $g$, which is related to the DWF by

\begin{equation*}
    g = \sqrt{\frac{3\pi c^3 \xi}{2 \tau \omega^2 n^2 V}}
\end{equation*}
where $\tau$ is the lifetime of the emitter, $V$ is the cavity mode volume, and $\xi$ is the DWF \cite{RevModPhys.73.565}. The strong coupling regime has been demonstrated in systems like single atoms in Fabry-P\'{e}rot cavities, single quantum dots in solid-state cavities, and in superconducting Josephson junction systems \cite{kimble1998strong,peter2005exciton,mckeever2003experimental,wallraff2004strong}, however, each of these systems has at least one fundamental drawback. Atom-Fabry-P\'{e}rot systems can be difficult to scale up to the size that more complex quantum information systems require. Quantum dots, as mesoscopic objects, can have a relatively wide distribution of emission wavelengths even when fabricated by the same process. Superconducting Josephson junction systems operate at millimeter wavelengths which are not compatible with distributed QIP.

While much progress has been made to make up for these various difficulties, SiC color centers in nanophotonic cavities inherently address all these issues. Firstly, color centers are scalable via integration into devices, as detailed in Section 3. Secondly, the linewidths of zero-phonon lines of color centers are generally much smaller than those of quantum dots. Finally, the near-infrared emission wavelengths provide some flexibility in the fabrication of scalable cavity arrays, since the features do not need to be as small as in the case of diamond emitters. This is not to imply that bringing color center systems into strong coupling CQED systems is devoid of challenges. The current state-of-the-art in color center-based nanophotonics is stymied by the low DWF and the presence of non-radiative recombination, which adds difficulty to achieving strong light-matter interaction in such systems. With careful nanodevice design, it is possible to modify $\xi$, $\tau$, and $V$ as discussed in Section 3. However, there has also been significant work in investigating multiple color centers acting as an ensemble, which scales the light-matter interaction strength by $\sqrt{N}$, where $N$ is the number of emitters in the ensemble, to achieve strong coupling and demonstrate polaritonic physics \cite{radulaski2017nonclassical, trivedi2019photon, majety2021quantum, patton2021all}.

\subsection{Measurement-Based Quantum Computing}
\begin{figure}[ht!]
    \centering
    \includegraphics [scale = 0.55]{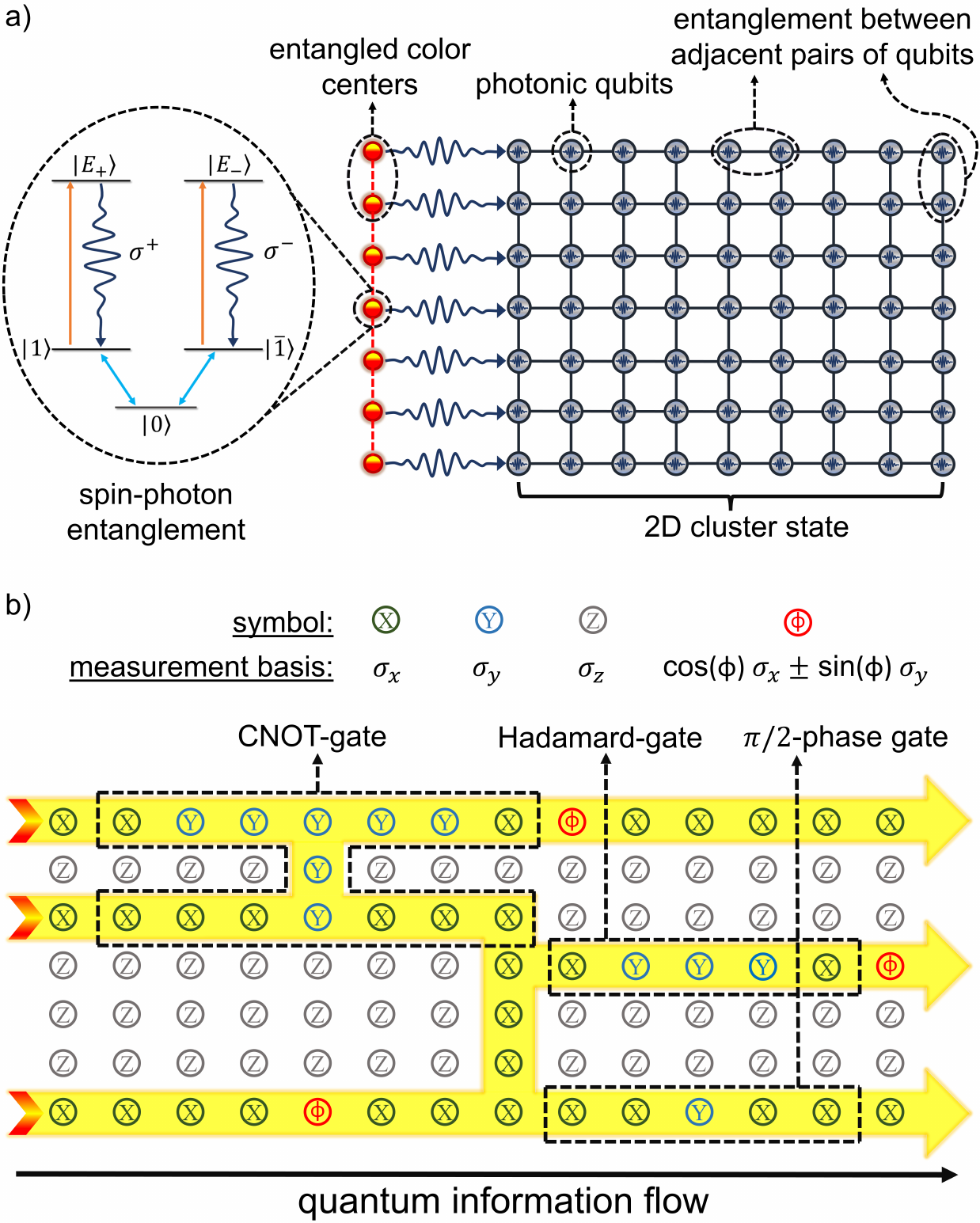}
    \caption{Color center implementation of all optical quantum computation with cluster states. a) Generation of the photonic cluster state using repeated emissions from entangled color centers. Inset figure shows the so called I-I energy level structure for spin-photon entanglement creation, exhibited in spin-$\frac{3}{2}$ and spin-1 emitters in SiC, such as silicon vacancy, divacancy, and NV center \cite{sophia2016spin}. b) Quantum logic network implemented on the cluster state via single qubit measurements in different bases \cite{raussendorf2001one,raussendorf2003two}. Operating principles are detailed in the main text.
    } 
    \label{fig:fig6}
    %\vspace{-0.5pt}
\end{figure}

One of the major goals of QIP is building a fully scalable quantum computer that can execute arbitrary quantum algorithms and distribute complex computations. With this motivation, the measurement-based quantum computer (MQC) \cite{raussendorf2001one,raussendorf2003two}, also known as  the one-way quantum computer, was proposed as an alternative to the more established circuit model. The universal resource for MQC is a highly entangled state known as the cluster state (Figure \ref{fig:fig6}a). Any sequential quantum circuit can be imprinted on the cluster state with only single qubit measurements. Standard procedure for cluster state generation is divided into two steps: (i) prepare all qubits in the $\ket{+} = (\ket{0}+\ket{1})/\sqrt{2}$ state (in the Pauli Z computational basis), and (ii) implement Ising-type next-neighbor interaction by applying the control-Z (entangling) gates between each pair of qubits. With each computational step, the entanglement is consumed by the subsequent measurement of qubits in a certain direction. Cluster state provides MQC with multiple advantages over the circuit-based model including decomposing large computational space into steps by re-utilizing the cluster \cite{raussendorf2001one}, concatenation of gate simulations \cite{raussendorf2003two}, distributed \cite{lim2006repeat, beals2013efficient, van2016path}, loss-tolerant\cite{varnava2006loss,zhan2020deterministic}, and fault-tolerant \cite{nielsen2005fault,raussendorf2007topological} quantum computation. Apart from MQC, the cluster state remains an efficient resource for quantum repeaters, as discussed previously.

Due to long coherence times and ease of manipulation of photons, optical quantum computing with cluster states \cite{nielsen2004optical} emerged from the combination of linear optics quantum computation (popularly known as KLM scheme) \cite{klm2001scheme} and MQC \cite{raussendorf2001one}. Using photons to produce highly entangled graph states is difficult, as photons naturally do not interact with each other. In earlier times, photonic graph states were generated either by measurement-induced entanglement \cite{bose1999proposal} (with cavities and multiplexer \cite{benjamin2009prospects}) or by spontaneous parametric down conversion (SPDC) combined with the so called fusion gates \cite{browne2005resource}. As their generation depended on probabilistic photodetection process, photonic graph states have reached experimental size of only ten photons \cite{wang2016ten_exp} with this approach. Deterministic generation of 1D photonic cluster strings was first realized by the Lindner-Rudolph protocol \cite{lindner2009proposal} from quantum dots with specific level structure and selection rules. Although 1D graph state suffices for quantum repeaters, the universal quantum computation requires a 2D graph state. This need was met in a proposal suggesting the use of entangled photons from entangled quantum emitters \cite{economou2010optically}. Basing on this idea, subsequent protocols \cite{sophia2017deterministic,sophia2018photonic,sophia2019deterministic,sophia2019generation} have been developed for generating arbitrarily large photonic graph states from solid-state emitters such as trapped ions, quantum dots, and NV centers in diamond. The protocol for 2D cluster state generation from NV centers in diamond \cite{sophia2019generation} requires only microwave pulses (for emitter manipulation), optical pumping (for photon generation), and waveplate and polarization photon counters (for measurement) to implement MQC, thus making it suitable for implementation with SiC color center photonics.

Although the silicon carbide protocols have been studied only in terms of 1D photonic cluster state generation \cite{sophia2016spin}, the field can benefit from the wide pool of studies in diamond NV center with whom divacancies and NV centers in SiC share similar electronic structure and selection rules. As a result, protocol for 2D photonic cluster state generation from NV centers in diamond \cite{sophia2019generation} can be directly implemented in divacancies and NV centers in SiC. Figure \ref{fig:fig6} manifests a perspective on MQC with SiC color centers. First, a highly entangled $n \times m$ cluster state is generated from entangled color centers where $n$ is the number of color centers and $m$ is the number of pump cycles (Figure \ref{fig:fig6}a). In the inset, the energy level structures of divacancy and NV center in SiC are shown. With zero magnetic field, the ground $\ket{\pm 1}$ and excited $\ket{E_\pm}$ states are degenerate but do not mix. Each ground state couples to one excited state with spontaneous emission of light having opposite circular polarization ($\sigma^\pm$) which generates spin-photon entanglement, while the $\ket{0}$ acts as an auxiliary state. Individual color centers can be entangled with each other via protocols involving electron and nuclear spin entanglement \cite{sophia2019generation}. After 2D cluster state generation, a quantum logic network is implemented on the cluster state with single qubit measurements (Figure \ref{fig:fig6}b). A $\sigma_x$ measurement acts like a \emph{wire} between adjacent pairs of qubits, a $\sigma_y$ measurement removes the qubit from linear cluster but links the neighboring qubits which simultaneously makes it work like a \emph{wire}, and a $\sigma_z$ measurement removes the qubit from the rest of the cluster state. Measurements in the cos($\phi)\sigma_x$ $\pm$ sin($\phi)\sigma_y$ basis are performed in order to realize arbitrary rotations to compensate for the randomness in rotations due to Pauli observables. However, in a more general sense, the measurement basis depends on the preceding measurements, which imposes a temporal order for a particular quantum logic implementation in a certain direction \cite{raussendorf2001one,raussendorf2003two}.

To implement these MQC protocols in color centers, the main challenge is to apply entangling gates to generate the 2D cluster state and do at least one round of measurement before the spin decoheres. Assuming 30 $\mu$s time to apply entangling gates with high fidelity ($\gtrsim 99\%$), as seen in diamond NV center \cite{sophia2018photonic}, this puts the lower limit on the required coherence time to hundreds of microseconds. This timescale makes divacancy and silicon vacancy in SiC currently the best suited candidates for SiC-based MQC, as they have spin coherence times in the milliseconds-seconds range \cite{seo2016quantum,babin2021fabrication} (see Table \ref{tab:Table1}). Experimental challenges such as spontaneous emission rate and inefficient photon collection due to low ZPL emission, and photon loss during multi-emitter entanglement also impede deterministic cluster state generation from realization in color centers \cite{sophia2016spin,sophia2017deterministic,sophia2018photonic,sophia2019generation}. These issues can be resolved by coupling SiC color centers to a photonic crystal cavity that extends into a single-mode waveguide attached to a detector (see Figure \ref{fig:fig2}a), fiber or an efficient grating coupler  \cite{bracher2017spontaneous, lukin20204h,  majety2021quantum, babin2021fabrication}. Therefore, integrated SiC photonics has the potential to be a rich toolbox for full-fledged execution of MQC.

\section{Summary}
In this perspective paper, we have presented integrated silicon carbide photonics as a powerful playground for quantum technology development. Although SiC has benefits as a material platform for QIP applications like CMOS compatible, wafer-scale nanofabrication processes, and availability of color centers with emission in telecommunications wavelength, progress needs to be made on the realm of entanglement generation in integrated color center devices.

In an international and interdisciplinary scientific effort, this field is moving to demonstrate unique capabilities in materials engineering, spin-photon entanglement, and cavity-emitter interaction. The potential impact on quantum networking, simulation, and computing has been drawing a larger community to explore SiC color centers, and we hope this paper will serve as a guiding material to connect concepts from related fields.

\section*{Acknowledgements}
This work is supported by the National Science Foundation (CAREER-2047564) and the UC Davis Summer GSR Award for Engineering or Computer-related Applications and Methods. 

\section*{Author Declarations}
\subsection*{Conflict of Interest}
The authors have no conflicts to disclose.

\subsection*{Data Availability}
Data sharing is not applicable to this article as no new data were created or analyzed in this study.

\bibliographystyle{unsrt}
\bibliography{biblio.bib}
\end{document}